\input phyzzx.tex

\tolerance=100
0
\voffset=-0.0cm

\hoffset=0.7cm

\sequentialequations

\def\rl{\rightline}

\def\t1{{\tilde 1}}

\def\t{\theta}

\REF{\LIN}{A. D. Linde, Phys. Lett. {\bf B108} (1982) 389; Phys. Lett. {\bf B114} (1982) 431.}

\REF{\STE}{A. Albrecht and P. J. Steinhardt, Phys. Rev. Lett. {\bf 48} (1982) 120.}

\REF{\ISS}{K. Intriligator, N. Seiberg and D. Shih, JHEP {\bf 0604} (2006) 021, [arXiv:hep-th/0602239].}

\REF{\DYN}{K. Intriligator and N. Seiberg, Class. Quant. Grav. {\bf 24} (2007) S741, [arXiv:hep-th/0702069].}

\REF{\KEN}{K. Intriligator, N. Seiberg and D. Shih, JHEP {\bf 0707} (2007) 017, [arXiv:hep-th/0703281].}

\REF{\RAY}{S. Ray, Phys. Lett {\bf B642} (2006) 13, [arXiv:hep-th/0607172]; arXiv:0708.2200[hep-th].}

\REF{\KOM}{Z. Komargodski and D. Shih, JHEP {\bf 0904} (2009) 093, arXiv:0902.0030[hep-th].}

\REF{\WEI}{S. R. Coleman and E. J. Weinberg, Phys. Rev. {\bf D7} (1973) 1888.}

\REF{\NEL}{A. Nelson and N. Seiberg, Nucl. Phys. {\bf B416} (1994) 46, [arXiv:hep-ph/9309299].}

\REF{\GUT}{A. H. Guth, Phys. Rev. {\bf D23} (1981) 347.}

\REF{\PRO}{S. W. Hawking, I. G. Moss and J. M. Stewart, Phys. Rev. {\bf D26} (1982) 2681; A. H. Guth and E. J. Weinberg, Nucl. Phys. {\bf B212} (1983) 321.}

\REF{\HYB}{A. D. Linde, Phys. Lett. {\bf B259} (1991) 8; Phys. Rev. {\bf D49} (1994) 748, [arXiv:astro-ph/9307002].}

\REF{\RIO}{A. Linde and A Riotto, Phys. Rev. {\bf D56} (1997) 1841, [arXiv:hep-ph/9703209].}

\REF{\COL}{S. R. Coleman, Phys. Rev. {\bf D15} (1977) 2929; S. R. Coleman and D. de Lucia, Phys. Rev. {\bf D21} (1980) 3305.}

\REF{\SHI}{D. Shih, JHEP {\bf 0802} (2008) 091, [arXiv:hep-th/0703196].}







\REF{\DINF}{E. Halyo, Phys. Lett. {\bf B387}  (1996) 43, [arXiv:hep-ph/9606423]; P. Binetruy and G. Dvali, Phys. Lett. {\bf B388}  (1996) 241, [arXiv:hep-ph/9606342].}

\REF{\HIL}{L. Boubekeur and D. H. Lyth, JCAP {\bf 0507} (2005) 010, [arXiv:hep-ph/0502047].}

\REF{\NEW}{E. Halyo, work in progress.}

\REF{\GEO}{C. Vafa, Journ. Math. Phys. {\bf 42} (2001) [arXiv:hep-th/0008142]; F. Cachazo, K. Intriligator and C. Vafa, Nucl. Phys. {\bf B603} (2001) 3, [arXiv:hep-th/0103067].}

\REF{\EDI}{E. Halyo, JHEP {\bf 0407} (2004) 080, [arXiv:hep-th/0312042]; [arXiv:hep-th/0402155]; [arXiv:hep-th/0405269].}

\REF{\REN}{R. Kallosh and A. Linde, JCAP {\bf 0310} (2003) 008, [arXiv:hep-th/0306058].}

\REF{\KIB}{M. B. Hindmarsh and T. W. B. Kibble, Rep. Prog. Phys. {\bf 58} (1995) 477, [arXiv:hep-ph/9411342].}

\REF{\COS}{E. Halyo, JHEP {\bf 0403} (2004) 047, [arXiv:hep-th/0312268]; arXiv:0906.2587[hep-th].}

\singlespace

\rl{SU-ITP-10/2}




\pagenumber=0

\normalspace

\medskip

\bigskip

\titlestyle{\bf{Inflation in Wess--Zumino Models}}

\smallskip

\author{ Edi Halyo{\footnote*{e--mail address: halyo@stanford.edu}}}

\smallskip

\centerline {Department of Physics} 

\centerline{Stanford University} 

\centerline {Stanford, CA 94305}

\smallskip

\vskip 2 cm

\titlestyle{\bf ABSTRACT}

We show that a class of Wess--Zumino models lead to inflation in supersymmetry and supergravity. This is due to the 
existence of a classically flat direction generic to these models. 
The pseudomodulus that parametrizes this flat direction
is the inflaton and obtains a small mass due to either one--loop or supergravity corrections giving rise to slow--roll 
inflation. 
At the end of inflation, the fields roll to a supersymmetric vacuum that arises from explicit R symmetry 
breaking.

\singlespace

\vskip 0.5cm

\endpage

\normalspace

\centerline{\bf 1. Introduction}

\medskip

A necessary condition for slow--roll inflation[\LIN,\STE] is the presence of a field, i.e. the inflaton, which correponds
to an almost flat direction of the scalar potential. 
Supersymmetric models generically have subspaces of field space with 
nonzero energy and classically flat directions 
parametrized by pseudomoduli. {\footnote1{Note that these are not real 
moduli since the models of inflation have isolated vacua with fixed VEVs.}} 
In such states, supersymmetry is necessarily 
broken and as a result, the pseudomoduli obtain potentials due to one--loop quantum effects. These potentials are 
logarithmic, giving rise 
to small masses for the pseudomoduli and causing them to roll slowly towards supersymmetric vacua.
Thus, the pseudomoduli play the role of the inflaton and inflation is naturally realized in 
supersymmetric models.

By now 
it is well--known that there are supersymmetric models with metastable nonsupersymmetric vacua[\ISS]. These models, which 
have been investigated for supersymmetry 
breaking purposes[\DYN,\KEN], also have pseudomoduli with classically flat 
potentials which are lifted due to one--loop effects in global supersymmetry or supergravity corrections. 
Therefore, 
they are naturally good candidates for models of inflation.
 An important example of this class are Wess--Zumino models 
which are supersymmetric models with the most general renormalizable superpotentials subject to (discrete and/or continous) 
global symmetries. In fact, recent results on Wess--Zumino models may be used to show that they are ideally suited for 
inflation in supersymmetry and
 supergravity. In this paper we show that, in both cases, Wess--Zumino models lead to F--term
inflation.

In the context of supersymmetry breaking, the metastable vacua in Wess--Zumino models have to be classically and
quantum mechanically stable[\ISS,\DYN,\KEN]. On the other hand, in order to exit inflation with vanishing vacuum energy, 
the fields have to relax to supersymmetric 
vacua. Thus, for a successful end to inflation, we have 
to demand exactly the opposite, that the models be classically 
unstable along the trajectory of the inflaton.

This paper is organized as follows. In section 2, we briefly review the 
general results on Wess--Zumino models that are relevant for inflation. We then describe a Wess--Zumino
model that realizes 
inflation in supersymmetry. In section 3, we show that the same model can give rise to inflation in supergravity. 
Section 4 inlcudes ideas on how to realize 
Wess--Zumino inflation in string theory, a discussion of our results and our 
conclusions.

\bigskip

\centerline{\bf 2. Wess--Zumino Inflation in Supersymmetry}

\medskip

Wess--Zumino models have been studied in great detail for supersymmetry breaking purposes[\DYN,\KEN,\RAY]. The new 
insights gained from these studies show that Wess--Zumino 
models can lead to inflation in supersymmetry and supergravity. 
Below, we summarize their properties that are important for inflation. 
Consider a generic Wess--Zumino model with chiral 
superfields $\phi_i$ and the (renormalizable) superpotential
$$W=F_i \phi_i+{1 \over 2} m_{ij} \phi_i \phi_j + {1 \over 6} \lambda_{ijk} \phi_i \phi_j \phi_k \eqno(1)$$
We assume that all fields have canonical Kahler potentials. In ref. [\RAY,\KOM] it was shown that, in a supersymmetry 
breaking vacuum, there is a classically flat direction, 
i.e. a pseudomodulus. In this vacuum all other fields are 
stabilized and get VEVs. One can perform a unitary transformation and redefine the fields so that the superpotential 
becomes
 $$W=X(F+{1 \over 2} \lambda_{ab} \phi_a \phi_b)+{1 \over 2} m_{ab} \phi_a \phi_b+{1 \over 6} \lambda_{abc} \phi_a \phi_b \phi_c \eqno(2)$$
where $X$ is the pseudomodulus that parametrizes the flat direction and $\phi_a$ are the transformed $\phi_i$ fields 
shifted by their VEVs. In this basis, the supersymmetry breaking vacuum is at
$\phi_a=0$ and arbitrary $X$. Since 
supersymmetry is broken, the bosonic masses of $\phi_a$ are shifted relative to those of their fermionic superpartners. 
As a result, there is a one--loop 
effective potential for $X$ generated by loops of $\phi_a$ and given by[\WEI]

$$V_1={1 \over {64 \pi^2}} STr{\cal M}^4 log{{\cal M}^2 \over \Lambda^2} \eqno(3)$$ 
where $STr$ is the supertrace operator, ${\cal M}$ is the (bosonic or fermionic) mass matrix and $\Lambda$ is a cutoff. 
We see that eq. (3) gives rise to a logarithmic potential and therefore
 a small mass for $X$. Thus, we conclude that the 
pseudomodulus $X$ is a good inflaton candidate. It has a classically flat potential which is lifted by one--loop effects 
that lead to 
a slow--roll towards the origin of field space (which is the metastable nonsupersymmetric vacuum).

In order to 
have a successful model of inflation, at the end of inflation the fields have to relax to a supersymmetric vacuum. However,
not all Wess--Zumino models preserve 
supersymmetry. It is well--known that, a sufficient condition for a generic model to 
have a supersymmetric vacuum is a superpotential that breaks R symmetry explicitly[\NEL]. 
Therefore, in order to have a 
supersymmetric vacuum, the superpotential given by eq. (2) has to break R symmetry. 
When both types of vacua are present, 
depending on the parameters of the model, the nonsupersymmetric vacuum may be classically or quantum mechanically stable.
Fortunately, it is quite 
easy to build Wess--Zumino models with superpotentials that break R symmetry explicitly.

In models 
of metastable supersymmetry breaking, one demands that there are no tachyonic (unstable) directions over the whole 
pseudomodulus space i.e. the complex line $X$. In ref. [\RAY] 
it was shown that this is equivalent to demanding that the 
matrix $m \lambda^{-1}$ be nilpotent. Moreover, 
one needs to make sure the matastable vacuum is also quantum mechanically 
stable, i.e. stable against tunneling. These demands for stability constrain the 
parameters such as 
$m_{ab}$ and $\lambda_{ab}$. On the other hand, for purposes of inflation, there is no need to demand the absence of 
unstable directions after inflation ends.
All that is required is a long enough slow--roll era that generates 60 e--folds 
and large enough scalar perturbations. In fact, since at the end inflation the fields have to reach 
a supersymmetric vacuum 
with vanishing energy, the requirement for a model of inflation is exactly the opposite; there has to be an 
unstable direction in field space either the trajectory of $X$. 
which leads to the supersymmetric vacuum. This will 
guarantee a successful exit from inflation. (Throughout the paper we sometimes refer to the origin of field space as the 
metastable 
vacuum even though it has to be unstable due to the above reasons.)

If there are no classical instabilities, 
an alternative way to exit inflation may be for the metastable vacuum to tunnel quickly to the supersymmetric one. 
This 
would lead to the old inflationary scenario[\GUT] with all the problems associated with it[\PRO]. Nevertheless, we will 
briefly consider this possibility in the following.

Consider the Wess--Zumino model with three chiral superfields $X, \phi_1,\phi_2$ and the superpotential[\RAY] 
(We assume that all fields have canonical Kahler potentials throughout the paper.)
$$W=-FX+\mu \phi_1 \phi_2+{1 \over 2} \lambda X \phi_1^2+{1 \over 6} g \phi_1 \phi_2^2 \eqno(4)$$
The scalar potential is given by
 $$V=|-F+{1 \over 2} \lambda \phi_1^2|^2+|\mu \phi_2+\lambda X \phi_1+{1 \over 6} g \phi_2^2|^2+|\mu \phi_1+{1 \over 3} g \phi_1 \phi_2|^2 \eqno(5)$$

For $\mu^2>\lambda F$ this model has a nonsupersymmetric vacuum at $\phi_1=\phi_2=0$ and arbitrary $X$. Thus, $X$ is a
pseudomodulus that parametrizes the flat direction and plays the role 
of the inflaton. As mentioned above, in order to 
exit inflation, the origin of field space has to be clasically unstable (i.e. it has to be a saddle point of $V$) which 
is the case for 
$\mu^2<\lambda F$. (In this range of parameters, the scalar potential in eq. (5) has another saddle point 
at $\phi_1=0$, $\phi_2=[-3 \mu \pm \sqrt{\mu^2-(2/3)g \lambda X \phi_1}]/g$ and 
arbitrary $X$, which we do not consider 
in the following.)

R symmetry is explicitly broken by the superpotential. With the charge assignments
$R[X]=R[\phi_2]=2$ 
and $R[\phi_1]=0$ the last term in eq. (4) breaks the R symmetry. As a result, there are supersymmetric vacua[\NEL] given 
by
$$X={3 \over 2}{\mu^2 \over {g \sqrt{2 \lambda F}}} \qquad \phi_1=\pm \sqrt{{{2F} \over \lambda}} \qquad \phi_2=-{{3 \mu} \over g} \eqno(6)$$
As expected, when $g \to 0$, the supersymmetric vacuum in eq. (6) escapes to infinity and disappears.

We initially start with large values for all fields. More precisely, we assume that 
initially $X$ and $\phi_2$ have large 
values whereas $\lambda \phi_1<H$ so that $X$ does not have a large tree level mass. Even though this initial state in 
somewhat unnatural, 
it is similar to the ones assumed in models of hybrid inflation[\HYB]. In those models, initially 
the inflaton has a large value whereas the value of the trigger field is relatively small 
for the same reason.
With this assumption, the massive fields $\phi_1$ and $\phi_2$ roll quickly to one of their extrema. Whether this is the 
origin of field space or the supersymmetric vacuum 
(or the second saddle point) obviously depends on the details of these 
initial values as well as properties of the field space. 
Therefore, answering this question requires a detailed analysis 
of the six dimensional field space and is quite complicated. 
(Note that when $\phi_1,\phi_2$ are 
in the supersymmetric vacuum given by the eq. (6), $X$ has a large mass, $m_X \sim H$, and inflation cannot take place. 
However, in the second saddle point 
$m_X$ vanishes at tree level since $\phi_1=0$ and this may lead to inflation.) 
Here we 
will simply assume that $\phi_1$ and $\phi_2$ roll towards the nonsupersymmetric metastable vacuum. Thus, the fields roll 
to $\phi_1=\phi_2=0$ and we can drop the last two terms 
in the superpotential.

The remaining superpotential contains only 
the pseudomodulus $X$ and leads to a constant scalar potantial. $X$ is the inflaton with a flat 
potential at tree level. 
In this vacuum, the bosonic masses squared are $$m_B^2={1 \over 2}[2 \mu^2+\lambda^2 |X|^2 -\epsilon  \lambda F \pm \sqrt{(\lambda^2 |X|^2 - \epsilon \lambda F)^2 +4 \mu^2 \lambda^2 |X|^2}] \eqno(7)$$
whereas the fermionic masses squared are

$$m_F^2={1 \over 2}[2 \mu^2+\lambda^2 |X|^2  \pm \sqrt{\lambda^2 |X|^4 +4 \mu^2 \lambda^2 |X|^2}] \eqno(8)$$
where $\epsilon=\pm 1$.
Since supersymmetry is broken ($F_X=-F$) loops of $\phi_1$ and $\phi_2$ generate a 
one--loop potential for $X$ given by eq. (3). With the masses in eqs. (7) and (8) and for $X>>\mu,F$ 
we get the total 
potential for $X$[\WEI]
$$V=V_0+V_1=F^2\left(1+{\lambda^2 \over {16 \pi^2}} log \left(X^2 \over \Lambda^2 \right)\right) \eqno(9)$$
where $\Lambda$ is a cutoff. The inflaton mass is $m_X=\lambda F/2 \sqrt{2} \pi X$ which is smaller than the Hubble 
constant $H=F/\sqrt{3}M_P$ for $X>\lambda M_P/2 \pi$. 
Inflation occurs when the inflaton $X$ slowly rolls down this 
potential and ends when one of the slow--roll conditions $|\epsilon|,|\eta|<<1$ is violated.

Since a successful end to 
inflation requires the origin to be classically unstable, we examine the bosonic masses in eq. (7) for $X>>\sqrt{F}>>\mu$. 
We find that out of the four masses 
squared, two are very large, $O(X^2)$, and one is very small and positive, 
$\sim 2 \mu^2F/\lambda X^2$. Finally, there is a tachyon with a very small negative mass squared
$\sim -2 \mu^2F/\lambda X^2$
which is the sign of the classical instability. For $\sqrt{F}>>\mu$ this tachyon mass is 
much smaller than the inflaton mass $\sim \lambda F/2 \pi X$. This means that as the inflaton 
slowly rolls down its 
potential during inflation, the tachyon sits at the top of the saddle point (the origin). In fact, the ratio of the 
tachyon mass to the inflaton mass is very small 
and independent of $X$, 
$m_{tach}/m_X \sim 4 \pi \mu/\sqrt{\lambda^3 F}<<1$. After inflation ends and $X$ rolls to smaller values, both the
tachyon and inflaton masses increase since they are both inversely proportional to $X$. Eventually, the tachyon mass 
becomes large as $X$ rolls towards the origin and the fields roll to the 
supersymmetric vacuum in eq. (6).
 Inflation fails
to occur for larger values of $\mu$, e.g. for $X>> \mu > \sqrt{F}$. In this case, the tachyon mass squared is 
$-F^2/X^2$ which is larger than the inflaton
 mass squared. This means that the fields roll to the supersymmetric vacuum 
before the inflaton has a chance to roll long enough to inflate the universe by 60 e--folds.

This scenario is basically 
F--term inflation previously considered in ref. [\RIO]. However, here we show that inflation occurs in a large class of 
models, i.e. Wess--Zumino models with 
R symmetry breaking superpotentials. The slow--roll conditions are satisfied if
$$\epsilon={M_P^2 \over 2} \left({V^{\prime} \over V}\right)^2={{2 \lambda^4 M_P^2} \over {(16 \pi^2)^2}}{1 \over X^2} \eqno(10)$$
and
$$\eta=M_P^2 \left(V^{''} \over V \right)=-{{\lambda^2 M_P^2} \over {8 \pi^2}}{1 \over X^2} \eqno(11)$$

We see that $|\epsilon|=\lambda^2 |\eta|/16 \pi^2$ is very small. Thus, $R=16 \epsilon$ is also very small and tensor
perturbations are negligible. In this model, inflation ends when 
$|\eta| \sim 1$. Inflating the universe by 60 e--folds requires

$$N={1 \over M_P^2} \int_{X_f}^{X_i} \left({V \over V^{\prime}}\right) dX={{4 \pi^2} \over {\lambda^2 M_P^2}} (X_i^2-X_f^2) \sim 60 \eqno(12)$$
The magnitude of scalar density perturbations is given by
$$A_s={H^2 \over {8 \pi^2 M_P^2 \epsilon}}={{16 \pi^2} \over {3 \lambda^4}} \left({F^2 \over M_P^4}\right)\left({X^2 \over M_P^2}\right) \sim 2\times 10^{-9} \eqno(13)$$
Using $X_i>>X_f$ and eq. (13)  we can obtain enough scalar density fluctuations with $F \sim 10^{-5} \lambda M_P^2$. 
Then, in order to get at least 60 e--folds we need 
$X_i > 1.3 \lambda M_P$. The condition on the spectral index of the 
scalar perturbations is 
$n_s=1-4\epsilon+2\eta \sim 0.96$ at $N \sim 60$. Since $|\epsilon|<<|\eta|$ we find from 
eqs. (11) and (12) that this condition can be satisfied with $\lambda \sim 0.1$.

Inflation ends when $|\eta| \sim 1$ 
which occurs around $X_f \sim (\lambda/2 \sqrt{2} \pi)M_P$. For $X<X_f$, the inflaton rolls down its potential quickly. 
When $X$ is small enough,
as described above, the tachyon mass becomes large and the fields roll to the supersymmetric 
vacuum.

Note that, even if $X$ gets close to the origin, there is no instability along the $X$ direction. Around 
$X \sim 0$, the one--loop potential becomes[\DYN,\RAY]

$$V_1=F^2+ {{\lambda^2 \mu^2 |X|^2} \over 32 \pi^2} F(x)+ O(|X|^4) \eqno(14)$$
where
$$F(x)={{1+x^2} \over x} log \left|{1+x} \over {1-x} \right| +2 log |1-x^2|-2 \eqno(15)$$
and $x=\lambda F/\mu^2 > 1$. Since $F(x)>0$, the point $X=0$ which is the endpoint of inflaton's trajectory is stable 
in the $X$ direction.

As mentioned above, if $\mu^2>\lambda F$ and there are no classical instabilities, the metastable 
vacuum may still decay to the supersymmetric one by tunneling.
The decay rate of the metastable vacuum is given by 
$\Gamma \sim e^{-S_I}$ where the action of the bounce is[\COL]
$$S_I={{27 \pi^2} \over 2}{(\Delta X)^4 \over \Delta V} \eqno(16)$$

Taking $\Delta X \sim \mu^2/\sqrt{F}$ and $\Delta V \sim F^2$ we find that $S_I \sim (27 \pi^2 /2) (\mu^2/F)^4$. If the 
origin is classically stable then $\mu^2>\lambda F$ and we find
the decay rate is greater than $\Gamma >e^{-130 \lambda^4}$.
For $\lambda<0.3$ the decay rate is very large and the false vacuum tunnels quickly to the supersymmetric vacuum. 
However, 
this leads to the old inflationary scenario[\GUT] with all its well--known problems[\PRO].

Using the above prescription, we can easily build many models of Wess--Zumino inflation. 
As another example, consider the model with the fields 
$X, \phi_1,\phi_{-1},\phi_3$ 
(where the subscripts on the fields denote their R charges) and the superpotential
$$W=FX+\lambda X \phi_1 \phi_{-1}- m_1\phi_1^2+m_2 \phi_{-1} \phi_3+m_3 \phi_3^2 \eqno(17)$$
 which was previously proposed (without the last term) as a model of supersymmetry breaking in ref. [\SHI]. 
We see that
with the choice $R[X]=2$, last term in eq. (17) breaks R symmetry explicitly. For $m_3=0$ the model has a 
nonsupersymmetric vacuum[\SHI]. When $m_3 \not =0$ 
a supersymmetric vacuum appears (in addition to the nonsupersymmetric 
one which becomes metastable). The scalar potential is given by
 $$V=|F+\lambda \phi_1 \phi_{-1}|^2+|\lambda X \phi_{-1}-2m_1 \phi_1|^2+|\lambda X \phi_1+m_2 \phi_3|^2+|m_2 \phi_{-1}+ 2m_3 \phi_3|^2 \eqno(18)$$
The nonsupersymmetric metastable vacuum is given by $\phi_1=\phi_{-1}=\phi_3=0 $ with
arbitrary $X$ at tree level. Thus, 
as before, $X$ is a pseudomodulus which parametrizes this flat direction and plays the role of the inflaton. 
Supersymmetry
is broken in this vacuum since $F_X=-F$ and, as a result, $X$ gets a potential (and mass) from one--loop effects.
This 
model also has a supersymmetric vacuum given by
$$\phi_1^2=-{F \over {2m_1}}X \qquad \phi_{-1}= {F \over {\lambda \phi_1}} \qquad \phi_3=-{m_2 \over {\lambda m_3}}{F \over {\lambda \phi_1}} \eqno(19)$$
where
$$X=\pm {m_2 \over \lambda} \sqrt{{m_1 \over m_3}} \eqno(20)$$

As expected the VEVs of $X, \phi_1, \phi_3$ are inversely proportional to $m_3$ which means that when $m_3 \to 0$ the 
supersymmetric vacuum escapes to infinity and disappears. 
If all the masses are of the same order of magnitude we find 
$\phi_1 \sim \phi_{-1} \sim \phi_3 \sim \sqrt{F/\lambda}$ and $X \sim m/\lambda$. We will not go into the details of 
inflation
 in this model since it leads to an inflationary scenario which is very similar to the one considered above.

\bigskip

\centerline{\bf 3. Wess--Zumino Inflation in Supergravity}

\medskip

The above scenario of Wess--Zumino inflation in global supersymmetry can be extended to supergravity. This is possible 
due to the special form of the $X$ dependent terms in the superpotential 
given by eq. (4). It is well--known that it is 
difficult to realize F--term inflation in supergravity due to the inflaton mass or $\eta$ problem (which lead to the 
advent of D--term 
inflation[\DINF] as an alternative). As shown in ref. [\RIO], this problem does 
not arise, if the 
inflaton superpotential has a special form (as in eq. (4) above). The supergravity scalar potential is given by
 $$V=e^K(|W_i+K_iW|^2-3|W|^2) \eqno(21)$$
 where 
the subscript $i$ denotes differentiation with respect to a field and $K$ is the Kahler potential which we take 
to be canonical for all fields. In eq. (21) we omitted all factors of $M_P$
which are easy to replace when necessary. 
Using eq. (4), the supergravity F--terms, $F_i=W_i+K_i W$ are given by
 $$F_X=-F+ \lambda \phi_1^2+{\bar X}(-F X+{1 \over 2} \lambda X \phi_1^2) \eqno(22)$$
 $$F_{\phi_1}=\mu \phi_2+ \lambda X \phi_1+ {g \over 6} \phi_2^2+ {\bar \phi_1}(-F X+ {1 \over 2} \lambda X \phi_1^2) \eqno(23)$$
and
$$F_{\phi_2}=\mu \phi_1+ {g \over 3} \phi_1 \phi_2+ {\bar \phi_2}(-F X+ {1 \over 2} \lambda X \phi_1^2) \eqno(24)$$

The scalar potential can be obtained by plugging eqs. (22)-(24) and (4) into eq. (21). First, we need to make sure that, 
as in the globally supersymmetric case, 
there are initial conditions that lead $\phi_1$ and $\phi_2$ to roll 
to the origin 
quickly. From the scalar potential it is easy to see that $m_X^2$ only gets corrections of order 
$O(\phi^2 F/M_P^2)$ or $O(F^2/M_P^2)$ where $\phi=\phi_1,\phi_2$.
For $\phi^2,F<<M_P^2$ these are about the same order 
of magnitude as the one--loop inflaton mass $\sim \lambda^2 F^2/8 \pi^2 X^2$.
 Thus, after taking supergravity corrections 
into account, $m_X$ is still small enough to lead to slow--roll inflation.{\footnote2{We show that
self--interactions of 
$X$ do not lead to a large mass term below.}} Moreover, it is easy to see that $\phi_1$ and $\phi_2$ have large masses as 
in the 
globally supersymmetric case.

In addition, we need to make sure that the origin of the field space remains a 
saddle point in supergravity. When $\phi_1=\phi_2=0$, $W_i=K_i=0$ and $W=-FX$, $\phi_1,\phi_2$ masses squared 
get a 
negative supergravity contribution of $\sim -F^2 X^2/M_P^4$. For the values of the parameters we used in the 
supersymmetric case, $F \sim 10^{-5}M_P^2$, $\lambda \sim 0.1$ and
$X \sim M_P/10$, we find that these supergravity 
contributions are much smaller than even the (small) tree level masses $ \sim 2 \mu^2 F/\lambda X^2$. Therefore, 
there is 
still one tachyon with a very small mass at the origin. This tachyon will sit at the top of the potential at the origin 
for a long time allowing slow--roll inflation 
in supergravity.

The potential for $X$ is obtained from eqs. (4) and (21)
$$\eqalignno{V_X&=e^{X{\bar X}}(|F+F {\bar X}X|^2-3|FX|^2) &(25)\cr
&=(1+X{\bar X}+{1\over 2}(X {\bar X})^2+ \ldots)(|F+F {\bar X}X|^2-3|FX|^2)\cr&=|F|^2(1+{1\over 2}(X {\bar X})^2 +\ldots)\cr}$$
where the dots indicate higher order terms that we neglect. 
We see that $X$ does not have a tree level mass but a quartic 
(and higher order) coupling. When $X$ has a large VEV this term 
induces a mass of $m_X^2 \sim X {\bar X} F^2/M_P^4$ 
(in addition to the one that arises from one--loop effects). The slow--roll of $X$ requires $m_X<H=F/\sqrt{3}M_P$ which 
means we need 
$X<\sqrt{3} M_P$ which is easily satisfied. Another possible contribution to $m_X$ arises from 
nonrenormalizable terms in the Kahler potential which we assumed to be canonical at tree
level. The Kahler potential may 
include nonrenormalizable terms like
 $$V_K={c \over M_P^2} \int~d^4 \theta (X {\bar X})^2= c\left({F_X \over M_P}\right)^2 X {\bar X} \eqno(26)$$
which gives $m_X= \sqrt{c} F_X/M_P= \sqrt{c} F/M_P$. We can assume that $\sqrt{c} \sim 1/4\pi$ since $V_K$ arises from 
one--loop effects and then $m_X<H$.
Thus, $X$ obtains a small mass due to one--loop effects and/or supergravity corrections 
and therefore is a good inflaton candidate.

Inflation occurs as $X$ slowly rolls towards the origin as a result of either 
the one--loop or supergravity potentials in eqs. (9) and (25) respectively.
 The dynamics of $X$ is determined by the 
competition between the one--loop superpotential in eq. (9) and the supergravity quartic term in eq. (25). By comparing 
these, 
we find that the former dominates at late times, i.e. for $X<\sqrt{\lambda}M_P/3$. Before that era, for larger 
values of $X$, i.e. for 
$\sqrt{\lambda}M_P/3<X<\sqrt{3}M_P$, we need to take supergravity contributions into account. 
Following ref.[\RIO], 
the number of e--folds obtained at later times, the supersymmetric era dominated by the one--loop 
potential is $N_{susy} \sim 4/\lambda$. 
At earlier times when $X$ is larger, the supergravity potential contributes 
$N_{sugra} \sim 9/\lambda$ to give the total number of e--folds $N=N_{susy}+N_{sugra}=13/\lambda$. 
If $\lambda \sim 0.1$ 
as in the globally supersymmetric case, we find that inflation lasts for $N \sim 130$ and the last 40 e--folds are due 
to the one--loop potential. 
Inflation ends when the slow--roll condition is violated, i.e. $|\eta| \sim 1$. The discussion 
of the scalar density perturbations
 and the scalar spectral index is exactly as in ref. [\RIO] and will not be repeated 
here.

At the end of inflation, $X$ reaches a critical value for which the mass of the tachyon at the origin is large. 
Then, the fields quickly roll to the supersymmetric vacuum. 
In a supersymmetric vacuum the supergravity F--terms in 
eqs. (22)-(24) have to vanish. Again, these are given by the supersymmetric F--terms plus corrections that 
are suppressed 
by $\phi^2/M_P^2$. Even though it is hard to solve these constraints analytically, clearly there must be a supersymmetric 
vacuum close to the globally supersymmetric one. 
Unfortunately, this supersymmetic vacuum described by the vanishing 
F--terms in eqs. (22)-(24) does not have vanishing energy. In supergravity, a supersymmetric vacuum has vanishing energy 
only if the superpotential vanishes in the vacuum. From 
eq. (21) and the VEVs in eq. (6) we see that this is not the case. 
We can remedy this by adding a fine--tuned constant to the superpotential in eq. (4). This does not affect the results of
 the 
previous section, i.e. Wess--Zumino inflation global supersymmetry, but leads to a vanishing vacuum energy in 
supergravity.

\bigskip

\centerline{\bf 4. Conclusions and Discussion}

\medskip

In this paper we showed that a large class of 
Wess--Zumino models leads to inflation in both supersymmetry and supergravity. The main property of Wess--Zumino models 
that make this 
possible is the existence of a pseudomodulus with a classicaly flat potential, (i.e. the inflaton), which 
obtains a small mass either at one--loop order or in supergravity. Inflation occurs as 
the inflaton slowly rolls toward 
the origin of field space. During this era with a large $X$, the tachyon at the origin is stable due to its very small 
mass (which is much smaller than the 
inflaton mass). After the slow--roll era ends, the inflaton reaches a value for 
which the tachyon mass becomes large and the fields roll to the supersymmetric vacuum.

Thus, Wess--Zumino models realize 
F--term inflation in both supersymmtry and supergravity. An important difference between the above models and that of 
ref. [\RIO] is the nature 
of the instability that causes 
the fields to roll to the supersymmetric vacuum. In F--term 
inflation, which is a type of hybrid inflation, initially the origin is classically stable and becomes unstable only 
after 
the inflaton rolls below a critical value. Then, the trigger field becomes tachyonic and the fields roll to the 
supersymmetric vacuum. In Wess--Zumino models of the type we discussed above, 
on the other hand, there is a classical 
instability from the beginning signaled by the tachyonic direction at the origin of field space. However,
this tachyon 
has a very small ($X$ dependent) mass until inflation ends and the inflaton rolls to a small value. Eventually, when $X$ 
becomes small enough, the tachyon mass becomes large 
and the fields settle at the supersymmetric vacuum.

It is interesting
to note that, since the its mass is so small, the dynamics of the tachyon may also lead to hilltop inflation[\HIL]. Since 
the tachyon mass depends on $X$, this requires
 an era in which $X$ hs a large value and is slowly rolling. This is 
precisely the scenario we described above, i.e. slow--roll inflation for $X$. Thus, it seems that a model similar to the
 one 
in section 2 may lead to hilltop inflation[\NEW].

The main open question in the above scenario is our assumption that, at the beginning of inflation, the fields $\phi_1$ 
and $\phi_2$ quickly roll to the origin rather than to the 
supersymmetric vacuum (or the second saddle point). In order to 
justify this, one should analyze in detail the evolution of the fields in the full six dimensional field space which 
is 
hard to do analytically. 
However, it seems that there must be a nonnegligible region of the initial phase space from which 
$\phi_1$ and $\phi_2$ roll to the origin. Nevertheless, since the whole scenario 
depends on this assumption, it is 
important to justify our assumption by investigating the evolution of the model numerically.

We obtained Wess--Zumino 
inflation in supersymmetry and supergravity. The natural extension seems to be realization of this scenario in string 
theory. It is not clear how to 
obtain generic Wess--Zumino models in string theory. On the other hand, it is very easy to 
obtain the purely inflaton part of the model, i.e. a model with only the $X$ dependent terms 
in eq. (2). For example, 
consider a compactification with an $A_2$ type singularity fibered over the complex plane $C(x)$ and defined by
 $$uv=z(z-mx)(z-m(x-a)) \eqno(27)$$
We wrap one D5 brane on each one of the two nodes ($S^2$s) of the singularity. On the D5 brane world--volume, this gives 
rise to a field theory with the gauge group 
$U(1)_1 \times U(1)_2$ and a matter sector with two singlets 
$\Phi_1,\Phi_2$ 
and a pair of bifundamentals $Q_{12},Q_{21}$ with charges $(1,-1)$ and $(-1,1)$ respectively[\GEO]. 
The diagonal group 
$U(1)_d=[U(1)_1+U(1)_2]/2$ decouples leaving the gauge group 
$U(1)=[U(1)_1-U(1)_2]/2$. The superpotential is given 
by[\GEO]
 $$W=Q_{12}Q_{21}(\Phi_2-\Phi_1)+{1 \over 2} m \Phi_1^2-ma \Phi_2 \eqno(28)$$
At energies below $m \sim M_s$, $\Phi_1$ decouples and we are left with
$$W=\Phi_2 (Q_{12}Q_{21}-ma)  \eqno(29)$$
With the identification $\Phi_2=X$ and $F=ma$ we obtain the $X$ dependent terms in eq. (2) which are exactly those that 
give rise to F--term inflation in supersymmetry and supergravity as in
 ref. [\RIO]. (The difference is that $X$ couples 
to $Q_{12}Q_{21}$ instead of $\phi_1^2$ in eq. (4) but this is not consequential.) This is very similar to inflation 
models that were 
obtained on other singular spaces[\EDI] some of which were D--term inflation models. In fact, since 
these models have ${\cal N}=2$ supersymmetry, F ad D--terms are equivalent (by an
$SU(2)$ rotation of the theory) and F 
and D--term inflation scenarios describe the same physics[\REN]. 
We see that it is quite easy to obtain the minimal 
F--term inflation on D5 branes
 wrapped on singularities. It would be interesting to generalize this result and obtain 
generic Wess--Zumino models in string theory that lead to inflation.

An important difference between the Wess--Zumino 
models in eq. (2) and those that can be obtained on D5 branes is the existence of gauge symmetries in the latter. 
Usually 
fields that appear in Wess--Zumino models cary only global charges and the superpotential is the most general one subject 
to the global symmetries. However, in brane 
models such as the one above, there is generically an Abelian group for each 
node. (These can be decoupled by making their gauge couplings very small but this does not affect the
 superpotential.) 
It is clear that if the models have Abelian gauge symmetries these will be spontaneously broken at the end of inflation 
and cosmic strings will be produced[\KIB]. In the brane 
description, these are D3 branes wrapped on the same singularities 
and correspond to the strings that arise in F--term inflation[\COS]. It would be interesting to examine such a brane 
inflation
 scenario and find out its implications for cosmic string production.


\bigskip

\centerline{\bf Acknowledgements}

I would like to thank the Stanford Institute for Theoretical Physics for hospitality.

\vfill

\refout

\end

\bye